\documentclass{PoS}

\title{Contribution title}

\title{What Lattice QCD tell us about the Landau Gauge Infrared Propagators}

\ShortTitle{What Lattice QCD tell us about the Landau Gauge Infrared Propagators}

\author{\speaker{Orlando Oliveira}%
       \\
       Centro de F\'{\i}sica Computacional, Departamento de F\'{\i}sica, Universidade de Coimbra, 3004-516 Coimbra, Portugal\\
       and \\
       Departamento de F\'{\i}sica, Instituto Tecnol\'ogico de Aeron\'autica, 12.228-900, S\~ao Jos\'e dos Campos, SP, Brazil \\
       E-mail: \email{orlando@teor.fis.uc.pt}}

\author{Paulo J. Silva %
       \\
       Centro de F\'{\i}sica Computacional, Departamento de F\'{\i}sica, Universidade de Coimbra, 3004-516 Coimbra, Portugal\\
       E-mail: \email{psilva@teor.fis.uc.pt}}

\author{Pedro Bicudo %
        \\
       CFTP, Departamento de F\'{\i}sica, Instituto Superior T\'ecnico, Av. Rovisco Pais, 1049-001 Lisboa, Portugal\\
       E-mail: \email{bicudo@ist.utl.pt}}
       
\abstract{
The calculation of the Landau gauge gluon propagator performed in Coimbra using lattice QCD simulations
is reviewed. Particular attention is given to the behavior of the gluon propagator in the infrared region and the value of
$D(0)$. In the second part of the article, the modeling of the lattice data using massive type propagators and Gribov type
propagators is discussed. Four different mass scales are required to describe the propagator over the full range of
momenta accessed by the simulations discussed here. Furthermore, assuming a momentum dependent gluon mass, 
we sketch on its functional dependence.
}

\FullConference{The many faces of QCD\\
		November 2-5, 2010\\
		Gent Belgium}

\begin{document}

\section{Introduction and Motivation}

The investigation of the gluon propagator in the Landau gauge, 
with particular emphasis on the deep infrared region,
has been a hot topic in the recent years,
with many contributions coming from lattice QCD simulations and the solutions of the 
Dyson-Schwinger equations. 

The QCD simulations for studying the gluon propagator in the Landau gauge with the SU(3) and SU(2) gauge groups, 
typically, use the Wilson action, which approaches the continuous action up to corrections of
order $\mathcal{O}(a^2)$. Furthermore, 
the gauge procedure ends up satisfying the lattice version of the Landau gauge condition
$\partial \cdot A = 0$. Again, the lattice gauge fixing condition reproduces the continuum condition up to corrections of
order $\mathcal{O}(a^2)$. 

In order to access the deep infrared properties of the gluon propagator, simulations have been performed
on large volumes with a lattice spacing $a \sim 0.2$ fm or, equivalently, $a^{-1} \sim 1$ GeV which, for SU(3) simulations,
corresponds to a $\beta = 5.7$. 
From the point of view of the lattice simulations, such a small $\beta$ stands at the beginning of the perturbative
scaling window and, therefore, one should expect a relatively large contribution from the finite lattice spacing effects. 

In \cite{Leinweber98} the finite lattice size and anisotropic effects were studied using a simulation with $\beta = 6.0$,
i.e. $a = 0.10$ fm. Methods to minimize these artifacts were also investigated. 
These methods, the cylindrical and conic cuts, are now widely used. 
This study was repeated in \cite{Leinweber99} together with the investigation of finite volume effects by
performing simulations for $\beta = 6.0$ and $\beta = 6.2$, i.e. $a = 0.075$ fm. 
If the finite lattice spacing effects seem to be under control,
the finite volume effects show up mainly in the infrared region, 
with the zero momentum gluon propagator decreasing with the lattice volume.
Similar conclusions were obtained using larger lattices volumes - see, for example,
\cite{Bonnet01,Oliveira11,Oliveira09,Cucchieri08}. If this effect is clearly seen in many lattice simulations, 
the extrapolation to the infinite volume is not so unanimous. In particular, depending on how you extrapolate to the
infinite volume \cite{Oliveira05}, one gets $D(0) = 0$, as predicted in \cite{Gribov,Zwanziger91}, or $D(0) \ne 0$.
Generalizations of the Zwanziger work do not require necessarily a vanishing zero momentum gluon propagator
\cite{Dudal08,Dudal08a,Gracey10}. Lattice simulations at finite volume, modulus the extrapolation to infinite volume,
are in good agreement with such scenarios \cite{Dudal10}. On the other hand, looking at quarks models derived from QCD 
that incorporate the propagation of the gluon \cite{Costa11}, it seems that phenomenology does not care about the
value of $D(0)$. Given that the zero momentum gluon propagator is connected with gluon confinement, it is a fundamental
quantity and, in this sense, it would be nice to be able to have an estimate for $D(0)$ from lattice simulations 
free of lattice artifacts.

In section \ref{glue_prop}, we review the large volume simulations, i.e. having at least V = (5.8 fm)$^4$,
for the gluon propagator performed by the authors and discuss the possible values for $D(0)$.

Besides the infrared behavior of the gluon propagator, we would like also to discuss the mass scales in the gluon propagator.
The lagrangean of QCD has no mass scale. However, dimensional transmutation introduces a scale, $\Lambda_{QCD}$,
in association with the ultraviolet properties of the theory. $\Lambda_{QCD}$ is used to separate phenomena which
belong to the non-perturbative regime of the theory, i.e. where the relevant momenta are small compared to $\Lambda_{QCD}$,
or to the perturbative regime.

From the point of view of the gluon propagator, if one wants to parameterize its functional
form more mass scales are required, being it as an effective gluon mass \cite{Oliveira10a} or masses \cite{Frasca08}, 
the Gribov mass
parameter \cite{Gribov,Zwanziger91} 
or via the introduction of masses which are related with condensates required to define a
non-perturbative Green's function generator for QCD
\cite{Dudal08,Dudal08a,Gracey10}. For example, in \cite{Dudal10} the non-perturbative part of the lattice
gluon propagator was described by
\begin{equation}
   D(q^2) = \frac{q^2 + M^2_1}{q^4 + M^2_2 \, q^2 + M^4_3} \, ,
   \label{eq11}
\end{equation}
with the mass scales being $M^2_1 = 2.15(13)$ GeV$^2$, $M^2_2 = 0.337(47)$ GeV$^2$ and
$M^2_3 = 0.51$ GeV$^2$. 
The precise meaning of these mass scales depends on the theoretical scenario under discussion. 
The gluon propagator as given by equation (\ref{eq11})
was used in \cite{Dudal11} to compute the lowest mass glueball states within the so-called Refined Gribov-Zwanziger
theory. Furthermore, in \cite{Baulieu10,Dudal10b} a Gribov-type propagator 
\begin{equation}
   D(q^2) = \frac{q^2}{q^4 + M^4}
\end{equation}
was used to investigated the QCD operators which can correspond to relevant physical degrees of freedom,
i.e. those operators whose propagator have a K\"all\'en-Lehmann representation. 
The Gribov type of propagator was also explored in
\cite{Sorella10} to develop a theoretical scenario to explain gluon confinement. Further, non-perturbative solutions of
gluon-ghost Dyson-Schwinger equations also allow for a dynamical generated gluon mass \cite{Cornwall82,Aguilar08,Aguilar08a}
which, again, require other mass scales than $\Lambda_{QCD}$.

In section \ref{glue_prop_masses}, we try to define the minimum set of mass scales which can describe the 
lattice gluon propagator over the full range of momenta using two possible scenarios. 
For the two situations considered, we find that the propagator requires the introduction of four mass scales.
It  remains to see what are the relevance of these masses, if any, to physical phenomena.

\section{Gluon Propagator from Large Volume Simulations \label{glue_prop}}

\begin{figure}[t] 
   \centering
   \includegraphics[scale=0.4]{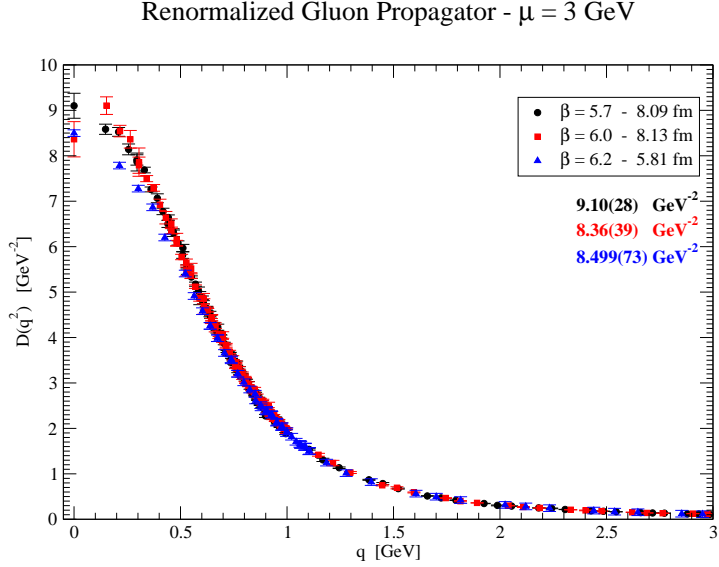} 
   \caption{Renormalized gluon propagator. The numbers in the picture are the values for $D(0)$ for the 
                 $\beta = 5.7$ simulation using a $44^4$ lattice, $\beta = 6.0$ with a $80^4$ lattice 
                 and $\beta = 6.2$ for a $80^4$ lattice.}
   \label{fig:glue}
\end{figure}

At Coimbra we have been computing the Landau gauge gluon propagator for several lattice spacings and volumes. 
Our goal is to perform a combined analysis of the finite lattice spacing and volume effects. We have just beginning such
an investigation and the results will be reported elsewhere. Here, we will focus on our large volume simulations.

Figure \ref{fig:glue} shows the renormalized gluon propagator for our largest lattice volumes for
several $\beta$ values. The corresponding values for $D(0)$ are also reported.
Definitions and the details on the computation of the propagator can be found in \cite{Silva04}.

\begin{figure}[t] 
   \centering
   \includegraphics[scale=0.4]{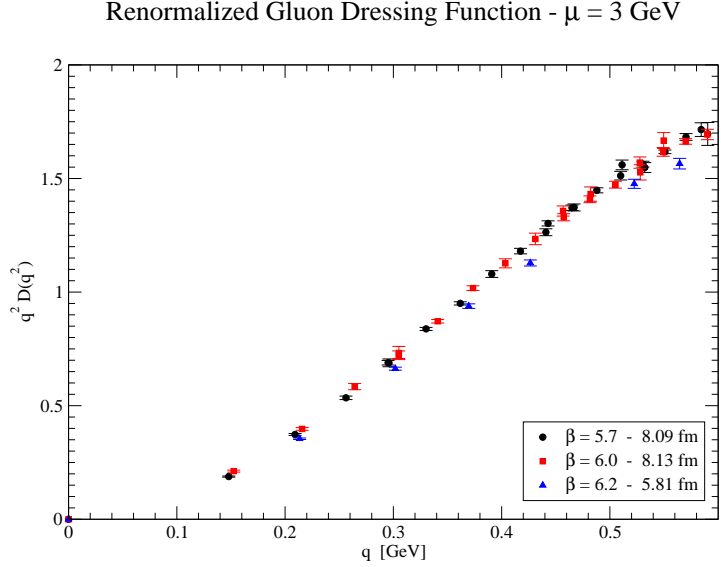} 
   \caption{Renormalized gluon dressing function.}
   \label{fig:gluedress}
\end{figure}

The propagators define a unique curve for momenta $q >  0.7$ GeV. For smaller momenta,
the propagator data for the finest lattice, which has the smallest physical volume, is below the corresponding data
points for the $\beta = 5.7$ and $\beta = 6.0$ simulations. 
This is certainly a finite volume effect which deserves to be further investigated. A closer look at the so called
gluon dressing function $q^2 D(q^2)$, see figure \ref{fig:gluedress}, shows small but finite volume effects.

For smaller lattice volumes, the gluon propagator follows the behavior observed in figure \ref{fig:glue}.
In all the simulations, no turnover on the gluon propagator at small momenta is observed. In what concerns the
value for $D(0)$, the values reported are, within two standard deviations, 
compatible for all simulations. A combined value, averaging
taking into account the statistical errors, gives $D(0) = 8.536(70)$ GeV$^{-2}$. This value is in good agreement
with the estimate of \cite{Dudal10} where $D(0) = 8.3(5)$ GeV$^{-2}$ and slightly above the result of 
the large volume simulations performed by the Berlin-Moscow-Adelaide group \cite{Bogolubsky09} where
$D(0) = 7.53(19)$ GeV$^{-2}$.

Can these results be an indication that $D(0)$ is finite and non-vanishing in the infinite volume limit? To our mind,
that depends on the way you want to look the gluon data. If one takes the propagator at its face value, then the
lattice simulations points clearly towards a finite and non-vanishing $D(0)$. However, one should not forget that
in the deep infrared region, one is relying on on-axis momenta 
and one is banging on the lattice walls.  Anyway, from the simulations performed so far,
the results look rather stable. However, if one looks at some details of the calculation, there are some holes
which should be investigated, in particular, the question of zero modes which can give a large contribution
to the deep infrared.

\begin{figure}[t]
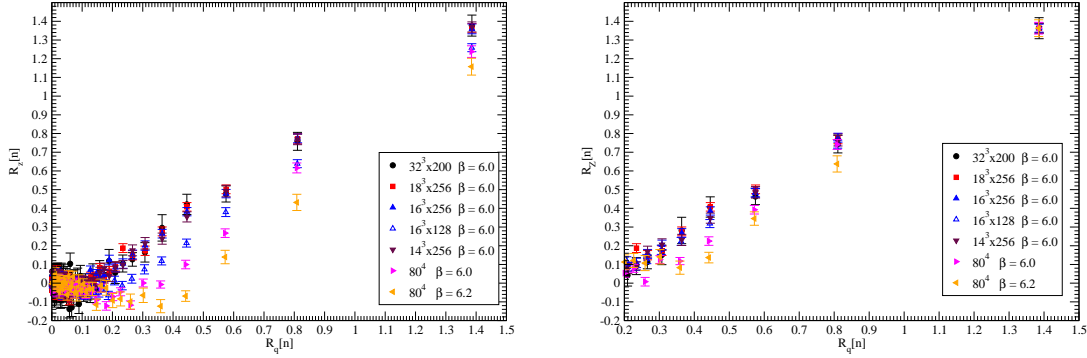
 
   \vspace{0.5cm}
   \centering
   \includegraphics[scale=0.28]{ratios_all.eps} \qquad
   \includegraphics[scale=0.28]{ratios_all_shifted.eps} \qquad
   \caption{Ratios of the gluon propagator. See \cite{Oliveira09} for details. In the left hand side, the plots shows
                 the ratios as directly obtained for the gluon propagator. In the right side, the ratios were shifted in order
                 to agree for the highest $R_q[n]$.}
   \label{fig:ratios}
\end{figure}

In \cite{Oliveira09} we have looked at the compatibility of the lattice gluon and ghost propagators with
a pure power law behavior in the deep infrared region using large asymmetric lattices. 
We will not reproduce the work here but will resume its main points. In the above cited work, we have investigated
ratios of propagators as a function of ratios of momenta. The ratios of propagators are renormalization
independent and, if the propagators are described by a pure power law, are universal functions.
Furthermore, the advantage of the ratios being that they suppress finite volume/lattice spacing
corrections. The results reported in \cite{Oliveira09} show that while the gluon propagator seems to follow
a pure power law in the deep infrared, the ghost propagator doesn't. Moreover, at least for the gluon propagator,
the results suggest a way to correct the propagator data. If this is taken into account as described in the appendix
of \cite{Oliveira09}, not only the corrected propagators become closer to each other, but also the 
corrected propagator decreases for momenta below 400 MeV. 

Figure \ref{fig:ratios} shows the ratios for the gluon propagator computed for several lattice simulations. In the left
hand side, the plot shows the ratios as measured from the gluon propagator. The differences between the different
data can be understood as due to finite volume effects. In the right hand side, the propagator data were shifted to
match the point at the highest $R_q[n]$, which corresponds to the smallest non-vanishing momenta available for
each lattice. From the right part of figure \ref{fig:ratios}, one can claim that the lattice propagator defines
a universal curve and, in this sense, claim that the gluon propagator is compatible with a pure power law in the
deep infrared. According to this analysis, the deep infrared propagator is described by
\begin{equation}
   D(q^2) = Z \left( q^2 \right)^{2 \kappa} \qquad \mbox{ with } \qquad \kappa = 0.528(14)
\end{equation}   
and $D(0)$ vanishes. Further details can be found in \cite{ManyFaces}. 

From the point of view of the phenomenology, the results of \cite{Costa11}, although obtained within a quark model, 
suggest that it would be quite hard to get some insight from phenomenology on the "correct" value for $D(0)$ and
distinguish, in this way, what value one should consider.

\section{Mass Scales from the Lattice Gluon Propagator \label{glue_prop_masses}}

\begin{figure}[t] 
   \centering
   \includegraphics[scale=0.4]{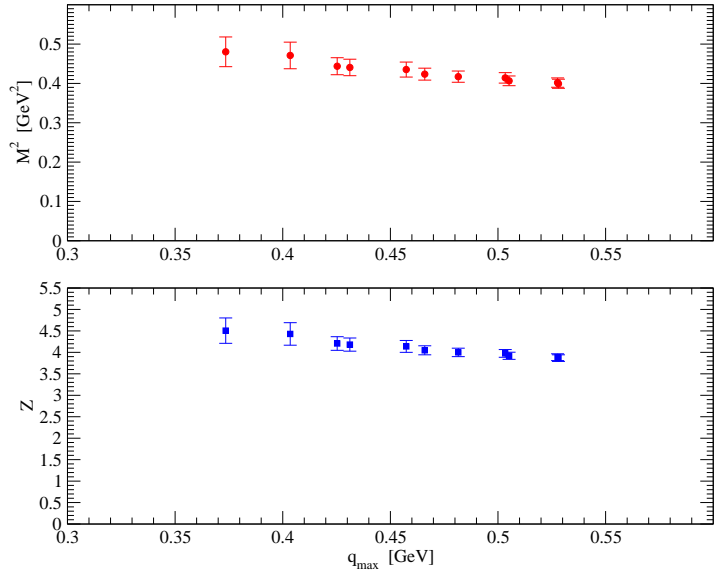} 
   \caption{The infrared gluon propagator as a massive type propagator.}
   \label{fig:massive_fit}
\end{figure}

We start our discussion of the mass scales for the
lattice gluon propagator assuming that it can be described by a massive type propagator \cite{Oliveira10a}, i.e. that
\begin{equation}
   D(q^2) = \frac{Z}{q^2 + M^2} \, .
   \label{massive_prop}
\end{equation}
The lattice data is fitted, in the range $q \in [ 0, q_{max} ]$, to equation (\ref{massive_prop}). Figure \ref{fig:massive_fit} 
resumes the outcome of fits.

From fitting the gluon propagator computed with the largest physical volume, i.e. the simulation using $\beta = 6.0$ and
an $80^4$ lattice, it follows that the lattice data can be
described by equation (\ref{massive_prop}) up to momenta $q = 482$ MeV, which has the smallest $\chi^2/d.o.f$,
with $M = 641(10)$ MeV and $Z = 3.99(13)$.
The corresponding $\chi^2/d.o.f.$ being 1.15. 
This parameters give a $D(0) = 10.58(41)$ GeV$^{-2}$, which is slightly above the
values claimed in the previous section.

The full range of momenta cannot be described by a massive type propagator, unless we allow $Z$ and $M$ to become
functions of the momenta. The computation of $Z(q^2)$ and $M^2(q^2)$ were investigated in \cite{Oliveira10a}. Performing
a slide window analysis, it was found that $Z(q^2)$ can be described by
\begin{equation}
  Z(q^2) = \frac{Z_0}{ \left[A + \ln\left( q^2 + m^2_0 \right) \right]^\gamma} \, ,
\end{equation}
where $\gamma$ is the anomalous gluon dimension, and the fits give 
$Z_0 \sim 1.0$, $A \sim -0.4$ and $m^2_0 \sim 1.57$ GeV$^2$ for a $\chi^2/d.o.f. = 1.8$. The mass squared is well
described for momenta above 1 GeV by
\begin{equation}
    M^2(q^2) = c_0 + c_1 \, q^2 \, \ln q^2 \, ,
    \label{massa_cornwall}
\end{equation}
with $q$ given in GeV, $c_0 \sim -0,3$ Ge$V^2$, $c_1 = -0.2$ for a $\chi^2/d.o.f. = 1.8$ and, for momenta below 1 GeV, is given by
\begin{equation}
    M^2(q^2) = 0.58(4) - 1.24(4) \, q^2  \, ;
    \label{massa_cornwall}
\end{equation}
the $\chi^2/d.o.f. = 1.8$. Note that (\ref{massa_cornwall}) is the expression of \cite{Cornwall82} for the dynamical generated gluon 
mass, where a non-perturbative solution of the gluon-ghost Dyson-Schwinger equations was investigated.

\begin{figure}[t] 
   \centering
   \includegraphics[scale=0.4]{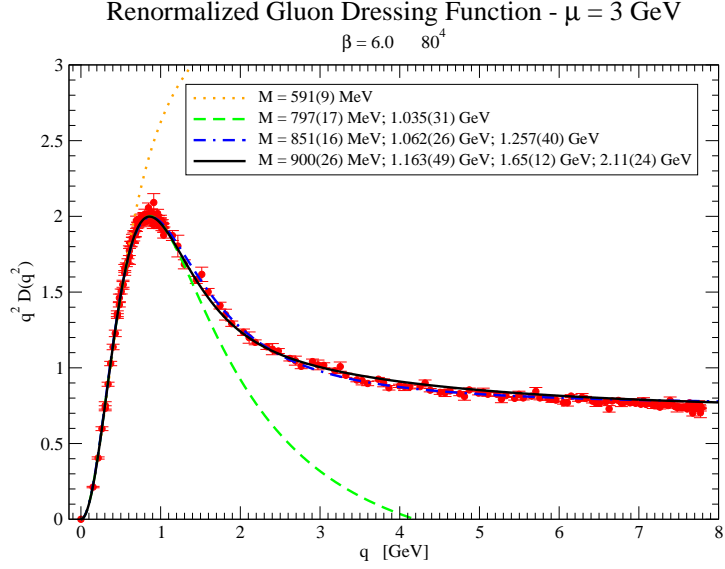} 
   \caption{The lattice gluon dressing function $q^2 D(q^2)$ as a sum of multi massive type propagators. 
                 The various curves correspond to various $N$. The legend includes the mass scales corresponding to
                 each fit.}
   \label{fig:multi_massive}
\end{figure}

Another possibility to describe the gluon propagator is to assume that $D(q^2)$ is given by a sum of massive type propagators,
i.e.
\begin{equation}
   D(q^2) = \sum^{N}_{i = 1}Ê\frac{Z_i}{q^2 + M^2_i} \, .
   \label{multi_massive}
\end{equation}
Note that  equation (\ref{multi_massive}) is a simplified K\"all\'en-Lehmann representation for the propagator. The figure shows that
by increasing $N$ one is able to described the lattice data over a wider range of momenta. The various fits give
\begin{eqnarray}
  & & 
            \Big[ 0.57  \Big( \{ 3.535(64), 0.5907(86) \} \Big) 1.4 \Big]  \nonumber \\
  &  & 
            \Big[ 1.52 \Big( \{ 17(3),  0.797(17) \}  \{ -17(3), 1.035(31) \} \Big) 1.5 \Big] \nonumber \\
  &   & 
            \Big[ 6.46  \Big( \{  31(6), 0.851(16) \}  \{ -52(11), 1.062(26) \}  \{ 22(9), 1.257(40) \} \Big)  1.6 \Big] \nonumber \\
  &    & 
            \Big[ 7.77 \Big(  \{  33(9), 0.900(26) \}  \{ -54(12), 1.163(49) \}  \{ 33(14), 1.65(12) \}  \{ -11(11), 2.11(24) \} \Big) 1.1 \Big]
            \nonumber
\end{eqnarray}            
where the results were written as $ \Big[ q_{max} \,  \Big( \{ Z_i, M_i \} , i = 1 \cdots N \Big) \, \chi^2/d.o.f.  \Big]$, with
$q$ and $M$ expressed in GeV. 
The last line
corresponds to the full range of momenta for the largest physical volume that we have simulated. We would like to call
the reader attention to the stability of the numbers, when increasing $N$ and the fitting range. Furthermore, note
that the lattice gluon propagator requires some negative $Z_i$. 
Given that (\ref{multi_massive})  can be viewed as a K\"all\'en-Lehmann representation,
negative $Z$ means that the gluon propagator is not an asymptotic state of the $S$-matrix. 

The above investigation shows that the lattice gluon propagator requires four mass scales. Note that the values for the 
masses are within typical hadronic masses.

\begin{figure}[t] 
   \centering
   \includegraphics[scale=0.4]{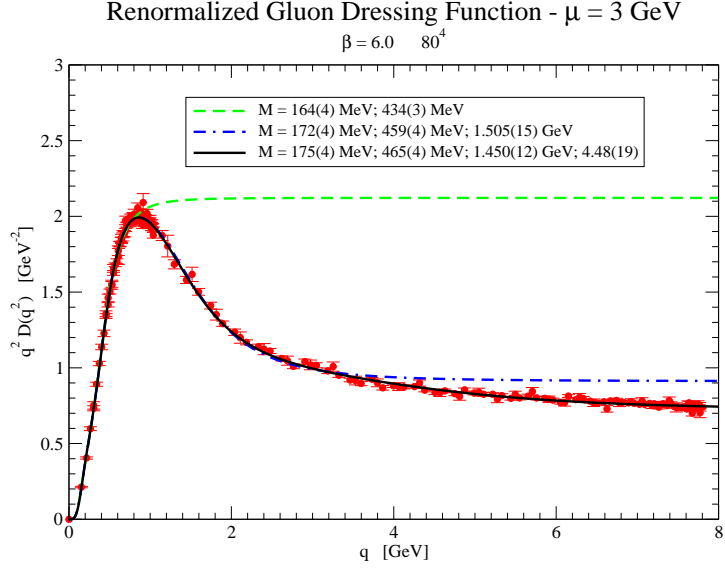} 
   \caption{The lattice gluon dressing function $q^2 D(q^2)$ as a sum of multi massive Gribov type propagators. 
                 The various curves correspond to various $N$. The legend includes the mass scales corresponding to
                 each fit.}
   \label{fig:multi_massive_gribov}
\end{figure}

As a final comment, we would like to address whether the gluon propagator can be described by a multiple mass
Gribov type formula, i.e.
\begin{equation}
   D(q^2) = \sum^{N}_{i = 1}ÊZ_i \, \frac{q^2}{q^4 + M^4_i} \, .
   \label{multi_massive_gribov}
\end{equation}
A similar investigation as for expression (\ref{multi_massive}) shows that when $N = 1$ the Gribov formula does not
match the lattice data, i.e. 
for $N = 1$ it turns out that $\chi^2/d.o.f.$ is always above 2. 
However, if one considers higher $N$, then the multiple mass Gribov type propagator
(\ref{multi_massive_gribov}) is compatible with the lattice data in an appropriate range of momenta. Following the same
notation as before, the fits give
\begin{eqnarray}
  &  & 
            \Big[ 0.90 \Big( \{ 0.430(19), 0.1639(38) \} , \{ 1.692(17), 0.4343(31) \} \Big) 1.6 \Big] \nonumber \\
  &   & 
            \Big[ 3.60  \Big( \{  0.487(19), 0.1724(36) \} , \{ 1.766(17) , 0.4588(37) \} , \{ -1.3412(83), 1.505(15) \} \Big) 1.6 \Big] \nonumber \\
  &    & 
            \Big[ 7.77 \Big(  \{  0.504(20), 0.1748(37) \} , \{ 1.773(17), 0.4647(43) \} ,  \nonumber \\
   &    &  \hspace{5.135cm}             \{ -1.306(12), 1.450(12) \} , \{ -0.251(12), 4.48(19) \} \Big) 1.3 \Big]
            \nonumber
\end{eqnarray}          
and can be seen in figure \ref{fig:multi_massive_gribov} together with the lattice data. Again, the description of the
lattice data over the full momentum range requires four mass scales.

\begin{figure}[t] 
   \centering
   \includegraphics[scale=0.4]{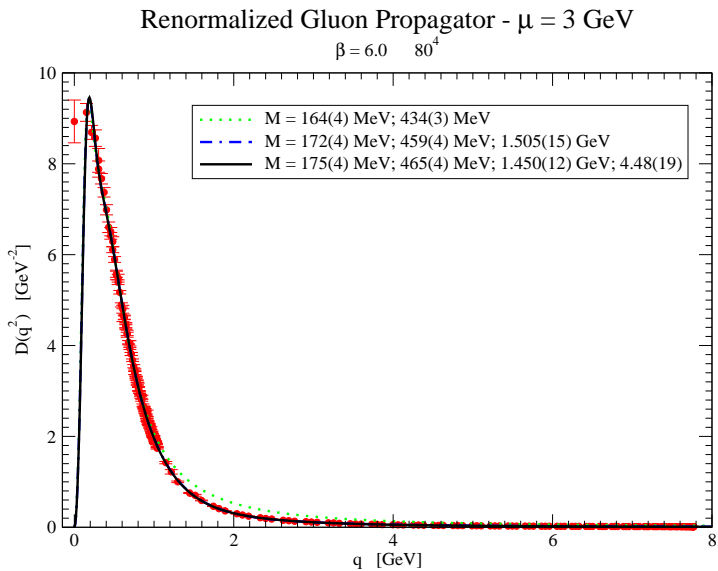} 
   \caption{The lattice gluon propagator as a sum of multi massive Gribov type propagators.}
   \label{fig:gmulti_massive_gribov}
\end{figure}

In what concerns the value for $D(0)$, the multiple massive type propagator gives a $D(0) \sim 10$ GeV$^{-2}$
(with large statistical errors),
while the multiple Gribov type propagator requires $D(0) = 0$. The Gribov-type propagator also provides an estimate where the
turn over can be observed. Indeed, the first derivative of $D(q^2)$ vanishes for $q = 190$ MeV, with the derivative
being negative for momenta above 0.19 GeV and positive below. The renormalized gluon propagator together with the
fit to (\ref{multi_massive_gribov}) can be seen in figure \ref{fig:gmulti_massive_gribov}.

\section{Final Comments}

In this article we review the calculation of the Landau gauge gluon propagator by OO and
PJS performed in Coimbra
using lattice QCD simulations. Although, the lattice data seems to point towards a finite and non-vanishing
$D(0)$ around 8 GeV$^{-2}$, there are still some open questions which deserve to be investigated. In particular, why
the ratios of propagators point towards a power law behavior in the infrared region and a $D(0) = 0$ needs to be understood.

In the second part of the article, the modeling of the lattice data using massive type propagators and Gribov type
propagators is discussed. It is shown that the full range of momenta can be described using four different mass scales. 
On the other side, assuming a mass dependent gluon mass, its functional dependence is sketched. 
In particular, it is shown that the lattice data is compatible with the parameterization used in the 
non-perturbative investigation of the Dyson-Schwinger equations. In what concerns the infrared
gluon propagator, we show that up to momenta around 500 MeV, the gluon propagator is well described by a massive type 
propagator, with an effective gluon mass around 640 MeV.

\section*{Acknowledgements}

The authors acknowledge financial support from FCT under contracts
PTDC/FIS/100968/2008, 
CERN/FP/109327/2009 and (P.J.S.) grant SFRH/BPD/40998/2007. 
O.O. acknowledges financial support from FAPESP.



\begin{thebibliography}{99}

\bibitem{Leinweber98} 
D. B. Leinweber, J. I. Skullerud, A. G. Williams, Claudio Parrinello, \textit{Phys. Rev.} \textbf{D58}, 031501 (1998).

\bibitem{Leinweber99} 
D. B. Leinweber, J. I. Skullerud, A. G. Williams, Claudio Parrinello, \textit{Phys. Rev.} \textbf{D60} , 94507 (1999); 
Erratum-ibid. \textbf{D61}, 079901 (2000).

\bibitem{Bonnet01}
F. D. R. Bonnet, P. O. Bowman, D. B. Leinweber, A. G. Williams, J. M. Zanotti,
\textit{Phys. Rev.} \textbf{D64}, 034501 (2001).

\bibitem{Oliveira11}
O. Oliveira, P. J. Silva, \textit{arXiv:1011.6107}; \textit{PoS} (\textbf{QCD-TNT09}), 033 (2009) [\textit{arXiv:0911.1643}].

\bibitem{Oliveira09}
O. Oliveira, P. J. Silva, \textit{Phys. Rev.} \textbf{D79}, 031501 (2009).

\bibitem{Cucchieri08}
A. Cucchieri, T. Mendes, \textit{Phys. Rev. Lett.} \textbf{100}, 241601 (2008)  [\textit{arXiv:0712.3517}].

\bibitem{Oliveira05}
O. Oliveira, P. J. Silva, \textit{PoS} (\textbf{LAT2005}), 287 (2005) [\textit{hep-lat/0509037}].

\bibitem{Gribov}
V. N. Gribov, \textit{Nucl. Phys.} \textbf{B139}, 1 (1978).

\bibitem{Zwanziger91}
D. Zwanziger, \textit{Nucl. Phys.} \textbf{B364}, 127 (1991).

\bibitem{Dudal08}
D. Dudal, S. P. Sorella, N. Vandersickel, H. Verschelde, \textit{Phys. Rev.} \textbf{D77}, 071501 (2008). 

\bibitem{Dudal08a}
D. Dudal, J. A. Gracey, S. P. Sorella, N. Vandersickel, H. Verschelde, \textit{Phys. Rev.} \textbf{D78}, 065047 (2008).

\bibitem{Gracey10}
J. A. Gracey, \textit{Phys. Rev.} \textbf{D82}, 085032 (2010).

\bibitem{Dudal10}
D. Dudal, O. Oliveira, N. Vandersickel, \textit{Phys. Rev.} \textbf{D81}, 074505 (2010).

\bibitem{Costa11}
P. Costa, O. Oliveira, P. J. Silva, \textit{Phys. Lett.} \textbf{B695}, 454 (2011).

\bibitem{Oliveira10a}
O. Oliveira, P. Bicudo, \textit{arXiv:1002.4151}; \textit{arXiv:1010.1975}.

\bibitem{Frasca08}
M. Frasca, \textit{Phys. Lett.} \textbf{B670} 73 (2008).

\bibitem{Dudal11}
D. Dudal, M. S. Guimaraes, S. P. Sorella, \textit{arXiv:1010.3638}.

\bibitem{Baulieu10}
L. Baulieu, D. Dudal, M. S. Guimaraes, M. Q. Huber, S. P. Sorella, N. Vandersickel, D. Zwanziger,
\textit{Phys. Rev.} \textbf{D82}, 025021 (2010) [\textit{arXiv:0912.5153}].

\bibitem{Dudal10b}
D. Dudal, N. Vandersickel, L. Baulieu, S. P. Sorella, M. S. Guimaraes, M. Q. Huber, O. Oliveira, D. Zwanziger,
[\textit{arXiv:1009.5846}].

\bibitem{Sorella10}
S. P. Sorella, \textit{arXiv:1006.4500}.

\bibitem{Cornwall82}
J. H. Cornwall, \textit{Phys. Rev.} \textbf{D26}, 1453 (1982).

\bibitem{Aguilar08}
A. C. Aguilar, J. Papavassiliou, \textit{Phys. Rev.} \textbf{D77}, 125022 (2008).

\bibitem{Aguilar08a}
A. C. Aguilar, D. Binosi, J. Papavassiliou, \textit{Phys. Rev.} \textbf{D78}, 025010 (2008).

\bibitem{Silva04}
P. J. Silva, O. Oliveira, \textit{Nucl. Phys.} \textbf{B690}, 177 (2004).

\bibitem{Bogolubsky09}
I. L. Bogolubsky, E.-M. Ilgenfritz, M.M\"uller-Preusker, A.Sternbeck,
\textit{Phys. Lett.} \textbf{B676}, 69 (2009)  [\textit{arXiv:0901.0736}].

\bibitem{Oliveira09}
O. Oliveira, P. J. Silva, \textit{Eur. Phys. J.} \textbf{C62}, 525 (2009).

\bibitem{ManyFaces}
Details can be found in \cite{Oliveira09} and on the transparencies of O. Oliveira talk on the webpage of the
Many Faces of QCD.

\end{thebibliography}
\end{document}